\documentclass[12pt]{article}

\usepackage{comment} 
\usepackage{amssymb}
\usepackage{color} 

\def \D {\hbox{d}}
\def \PIII {{\rm P_{\rm III}}} 
\def \PIIIDeight {{\rm P_{\rm III}-D_8}} 
\def\fba{U}

\def \ccomma{\raise 2pt\hbox{,}\ } 

\begin{document}

\title{On a dynamical system linked to the BKL scenario}

\author{Robert Conte$^{1,2}$
\\ $^1$ Universit\'e Paris-Saclay, ENS Paris-Saclay, CNRS,
\\      Centre Borelli, F-91190 Gif-sur-Yvette, France
\\ $^2$ Department of Mathematics, The University of Hong Kong,
\\      Pokfulam, Hong Kong.
\\
Robert.Conte@CEA.FR,       \\ ORCID https://orcid.org/0000-0002-1840-5095
}
\vspace{10pt}

\maketitle

\baselineskip=12truept

\begin{abstract} 
We consider the six-dimensional dynamical system in three components
introduced by Ryan to describe the scenario of Belinskii, Khalatnikov and Lifshitz 
to the cosmological singularity
when the spatial metric tensor is not diagonal.
Despite its nonintegrability, recently proven by Goldstein and Piechocki,
the three four-dimensional systems defined by canceling one of the three components
happen to be integrable.
We express their general solution as a rational function of, respectively,
two exponential functions, a third Painlev\'e function, two exponential functions. 
\end{abstract}

\medskip


\tableofcontents

\section{The dynamical system}
\label{sectionSummary}

The approach to the cosmological singularity
in the scenario of Belinskii, Khalatnikov and Lifshitz \cite{BLK,BKL}
is best described in the synchronous frame by the metric of Bianchi type IX, 
\begin{eqnarray} 
& &
\D s^2=\D t^2 - \gamma_{ab}(t) e_\alpha^a e_\beta^b \D x^\alpha \D x^\beta,
\label{eqds2}
\end{eqnarray} 
in which 
$t$ is the synchronous time,
$x^1,x^2,x^3$ the spatial coordinates,
$e^1,e^2,e^3$ the three frame vectors,
and $\gamma_{ab}(t)$ the spatial metric tensor,
independent of the spatial coordinates.
\smallskip
Indeed, the dynamics then reduces to a six-dimensional dynamical system
only involving the six components of $\gamma(t)$
(a real, symmetric, third order matrix).
However, this system is quite complicated \cite[Eqs.~(2.14)--(2.19)]{Ryan}
when written in the three eigenvalues $\Gamma_a$ of $\gamma$ 
and the three Euler angles of the rotation matrix
which diagonalizes $\gamma$.
At this stage, Ryan \cite{Ryan}
noticed that the approximation
\begin{eqnarray} 
& & \Gamma_1 \gg \Gamma_2 \gg \Gamma_3    
\end{eqnarray} 
makes the three Euler angles tend to constant values near the cosmological singularity,
thus defining an approximate
dynamical system in three components $(a,b,c)$
(essentially $(\Gamma_1, \Gamma_2, \Gamma_3)$)   
made of three second order ordinary differential equations,
\begin{eqnarray}
& &
\frac{\D^2 \log a}{\D t^2} = - a^2+ \frac{b}{a}\ccomma
\frac{\D^2 \log b}{\D t^2} = + a^2- \frac{b}{a}+\frac{c}{b}\ccomma
\frac{\D^2 \log c}{\D t^2} = + a^2- \frac{c}{b}\cdot
\label{System-abc}
\end{eqnarray} 
The validity of such an approximation has been investigated in detail by 
Parnovsky \cite{Parnovsky2022},
Goldstein and Piechocki \cite{Goldstein-Piechocki-2022},
G\'o\'zd\'z, P{\c{e}}drak and Piechocki \cite{GPP}.
For an additional review on the BKL scenario, we refer the reader to \cite{BKL1982}.

We rewrite (\ref{System-abc}) as an equivalent system in the three components $(a^2,b/a,c/b)$,
\begin{eqnarray}
& & {\hskip -15.0 truemm}
\begin{array}{ll}
\displaystyle{
\frac{\D^2 \log (a^2)}{\D t^2} = b/a - 2 a^2,
\frac{\D^2 \log (b/a)}{\D t^2} = 2 a^2 -2 b/a + c/b,
\frac{\D^2 \log (c/b)}{\D t^2} = b/a -2 c/b.
}
\end{array}
\label{System-natural}
\end{eqnarray} 
This system admits a first integral \cite[Eq.~(2.28)]{Ryan},
\begin{eqnarray}
& & 
K=\frac{\D \log b}{\D t} \frac{\D \log c}{\D t} 
 +\frac{\D \log c}{\D t} \frac{\D \log a}{\D t} 
 +\frac{\D \log a}{\D t} \frac{\D \log b}{\D t} 
 - a^2- \frac{b}{a}- \frac{c}{b}
\nonumber\\ & & \phantom{K}
= \frac{3}{4}\left(\frac{\D \log (a^2)}{\D t}\right)^2 
 +           \left(\frac{\D \log (b/a)}{\D t}\right)^2
 + 2\left(\frac{\D \log (a^2)}{\D t}\right) \left(\frac{\D \log (b/a)}{\D t}\right)
\nonumber\\ & & \phantom{K=}
 +  \left(\frac{\D \log (a^2)}{\D t}\right) \left(\frac{\D \log (c/b)}{\D t}\right)
 +  \left(\frac{\D \log (b/a)}{\D t}\right) \left(\frac{\D \log (c/b)}{\D t}\right)
\nonumber\\ & & \phantom{K=}
 - a^2- \frac{b}{a}- \frac{c}{b}.
\label{First-integral}
\end{eqnarray}
Although this first integral (a component of the Ricci tensor) must be set to zero in cosmology,
we keep it arbitrary here.

The invariance of (\ref{System-abc})
under the scaling transformation
\begin{eqnarray}
& &
(t,a,b,c,a^2,b/a,c/b) \to 
(\lambda^{-1} t,\lambda a,\lambda^3 b,\lambda^5 c, \lambda^2 a^2,\lambda^2 b/a,\lambda^2 c/b),
\end{eqnarray} 
led Goldstein and Piechocki \cite{Goldstein-Piechocki-2022}
to find the one-parameter solution (the arbitrary constant is the origin of $t$, omitted),
\begin{eqnarray}
& &
a= \pm \frac{3}{t}, 
b= \pm \frac{30}{t^3}, 
c= \pm \frac{120}{t^5}, 
\frac{a^2}{9}=\frac{b/a}{10}=\frac{c/b}{4}=\frac{1}{t^2},
\label{solution-scaled}
\end{eqnarray} 
easily extrapolated to one more parameter when one allows $K$ to be nonzero
\cite[p 216 col 2 line -5]{Goldstein-Piechocki-2022}, 
\begin{eqnarray}
& &
\frac{a^2}{9}=\frac{b/a}{10}=\frac{c/b}{4}=\frac{k^2}{\sinh^2 (k t)}, k^2=\frac{K}{23}\cdot
\label{solutions-trigo}
\end{eqnarray} 

When computing the Fuchs indices of the linearized system near the above 
one-parameter solution,
the same authors found the six values defined by 
\begin{eqnarray}
& &
(j+1)(j-2) \left[(j-1/2)^4 +95/2 (j-1/2)^2 +5569/16\right]=0,
\label{eqFuchs}
\end{eqnarray} 
The values $-1, 2$ correspond respectively to $t_0$ and $K$,
and the four others are not real.
This proves that the solution (\ref{solutions-trigo}) is the only singlevalued solution,
and this is 
a strong indication of the chaotic nature of the approach to the
cosmological singularity.

Our interest here is different,
and mainly motivated by mathematical similarities between the approximate system (\ref{System-abc})
and another six-dimensional Bianchi IX system,
defined by the assumption of a diagonal spatial metric tensor $\gamma(t)$.
In this so-called mixmaster model
\cite{LandauLifshitzTheorieChamps}
\cite[(3.5)--(3.6)]{BLK}
($(A,B,C)$ denote the components of the diagonal tensor $\gamma$),
\begin{equation}
(\log A)'' = A^2 - (B-C)^2,
(\log B)'' = B^2 - (C-A)^2,
(\log C)'' = C^2 - (A-B)^2,
\label{eqBianchiIXMixmaster}
\end{equation}
all singlevalued particular solutions but one \cite{LMC1994}
have been obtained explicitly.
Since this unknown missing solution
should depend on four arbitrary constants
and be an extrapolation of a three-parameter elliptic solution, 
we think it useful to examine the nature of the solutions of 
the three four-dimensional systems defined 
by canceling one of the three components in (\ref{System-natural}).
Since such a cancellation is unphysical
(because the volume $\det \gamma$ has been assumed nonzero when rescaling the coefficient of $\D t^2$
to unity in (\ref{eqds2})),
the obtained solutions will have no physical significance,
but their mathematical structure might be of interest.

The results are the following.
First, the general solutions of these three four-dimensional systems
are indeed singlevalued.
Second, two of them are mathema\-tically identical to two of
the four-parameter explicit solutions of the mixmaster.
However, as explained later,
the third one is of no help to tackle the missing mixmaster solution
because their analytic nature is different.
\medskip

\section{Four-dimensional cases of (\ref{System-natural})}
\label{section-Solutions-system-natural}

To cancel one of the three components $a^2,b/a,c/b$,
one first assigns it to a nonzero constant,
which cancels the left hand side,
then sets this constant to zero.
The system (\ref{System-natural}),  (\ref{First-integral}) then reduces to two equations
invariant by permutation of the two nonzero components
(the expression (\ref{First-integral}) reduces to zero).
The resulting system admits two first integrals and,
as we now see, 
a single valued general solution.

\subsection{Case $b/a=0$} 

This is the simplest case, since
the system 
\begin{eqnarray}
& & b/a=0,
\frac{\D^2 \log (a^2)}{\D t^2} = - 2 a^2,
\frac{\D^2 \log (c/b)}{\D t^2} = -2 c/b,
\label{System-natural-fba-zero}
\end{eqnarray} 
is uncoupled. 
Its four-parameter general solution
\begin{eqnarray}
& &
a^2=- \frac{k_a^2}{\sinh^2 k_a (t-t_a)},
b/a=0,
c/b=- \frac{k_c^2}{\sinh^2 k_c (t-t_c)},
\label{System-natural-fba-zero-sol}
\end{eqnarray} 
is identical to the general solution of the Bianchi IX mixmaster (\ref{eqBianchiIXMixmaster})
when two components are equal \cite{Taub},
the correspondence being
\begin{eqnarray}
& &
B=C,
A^2=\hbox{same as } a^2,
A B=\hbox{same as } c/b.
\end{eqnarray}

\subsection{Case $c/b=0$}

The system 
\begin{eqnarray}
& &
c/b=0,
\frac{\D^2 \log (a^2)}{\D t^2} = 2 b/a - 2 a^2,
\frac{\D^2 \log (b/a)}{\D t^2} = 2 a^2 - 2 b/a,
\label{System-natural-fcb-zero}
\end{eqnarray} 
admits two first integrals $K_1$ and $K_2$ defined by
\begin{eqnarray}
& & 2 K_1=\frac{\D \log (a^2)}{\D t} + \frac{\D \log (b/a)}{\D t},
    K_2^2=e^{-2 K_1 t} a^2 \frac{b}{a},
\end{eqnarray} 
and integrates with the third Painlev\'e function
(more precisely with $\PIIIDeight$ \cite{CMBook2})
\begin{eqnarray}
& &
\left\lbrace
\begin{array}{ll}
\displaystyle{
a^2(t)=-\frac{K_1^2 \alpha}{2} F(T),
\frac{b}{a}=- \frac{2 K_2^2 e^{2 K_1 t}}{K_1^2 \alpha F(T)},
T=- \frac{4 K_2^2 e^{2 K_1 t}}{K_1^4 \alpha\beta},
}\\ \displaystyle{
\frac{\D^2 F}{\D T^2} = \frac{1}{F} \left(\frac{\D F}{\D T}\right)^2 - \frac{\D F}{F \D T} 
+ \alpha F^2 + \frac{\beta}{4 T}+ \frac{\gamma F^3}{4 T^2}+ \frac{\delta}{4 F},
\alpha\beta\not=0,\gamma=\delta=0.
}
\end{array}
\right.
\end{eqnarray} 

Such a $\PIII$ function is also a solution 
of the Bianchi IX mixmaster (\ref{eqBianchiIXMixmaster})
in the case \cite{C2008B9P3}
\begin{eqnarray}
& &
A=0, (\log BC)'=\hbox{nonzero constant}, B=\PIII.
\end{eqnarray} 

\textbf{Subsection{ Case $a^2=0$}}
\subsection{Case $a^2=0$}

The last case is not so simple.
The system (\ref{System-natural}), which becomes
\begin{eqnarray}
& &
a^2=0,
\frac{\D^2 \log (b/a)}{\D t^2} = -2 b/a + c/b,
\frac{\D^2 \log (c/b)}{\D t^2} = b/a -2 c/b,
\label{System-natural-fa2-zero}
\end{eqnarray} 
admits two first integrals  (one even, one odd) polynomial in $b/a,c/b,(\log(b/a))',(\log(c/b))'$ 
\begin{eqnarray}
& &
\left\lbrace
\begin{array}{ll}
\displaystyle{ 	
K_a= 
             \left(\frac{\D \log (b/a)}{\D t}\right)^2
 +           \left(\frac{\D \log (c/b)}{\D t}\right)^2
 +  \left(\frac{\D \log (b/a)}{\D t}\right) \left(\frac{\D \log (c/b)}{\D t}\right)
 + 3 \frac{b}{a}+ 3 \frac{c}{b},
}\\ \displaystyle{
K_b=2 \left(\frac{\D \log (b/a)}{\D t}\right)^3 - 2 \left(\frac{\D \log (c/b)}{\D t}\right)^3
}\\ \displaystyle{
\phantom{1234}
 + 3       \frac{\D \log (b/a)}{\D t} \frac{\D \log (c/b)}{\D t}
     \left(\frac{\D \log (b/a)}{\D t}-\frac{\D \log (c/b)}{\D t}\right)
}\\ \displaystyle{
\phantom{1234}
		 -18 \left( (c/b) \frac{\D \log (b/a)}{\D t} - (b/a) \frac{\D \log (c/b)}{\D t}\right)
		 +9 \left(\frac{\D \log (b/a)}{\D t} - \frac{\D \log (c/b)}{\D t}\right).
}                                          
\end{array}
\right.
\end{eqnarray} 

The elimination of $c/b$ defines a single second order ODE for $b/a$,
\begin{eqnarray}
& &
\left\lbrace
\begin{array}{ll}
\displaystyle{ 
\fba=\frac{b}{a},
{\fba''}^3
+(K_a \fba - 9 \fba^2) {\fba''}^2
+(K_b \fba \fba' -3 K_a {\fba'}^2 -54 \fba {\fba'}^2) \fba''
}\\ \displaystyle{\phantom{123456}
- \frac{4 K_a^3}{27} \fba^3 
+ \frac{K_b^2}{27 }  \fba^3 + 4 K_a^2 \fba^4 -36 K_a \fba^5+108 \fba^6
+ 2 K_b \fba^3 \fba' 
}\\ \displaystyle{\phantom{123456}
+ K_a^2  \fba {\fba'}^2
- 18 K_a  \fba^2 {\fba'}^2
+ 108  \fba^3 {\fba'}^2
- K_b   {\fba'}^3
-27 {\fba'}^4
=0.
}
\end{array}
\right.
\label{eqode2fba}
\end{eqnarray}

Some two-parameter solutions of 
the equations (\ref{System-natural-fa2-zero}), (\ref{eqode2fba})
have already been found \cite[Eq.~(2.35)]{Ryan}
($t_0,k$ arbitrary),
\begin{eqnarray}
& & {\hskip -5.0truemm}
\left\lbrace
\begin{array}{ll}
\displaystyle{ 	
%
%
a^2=0, b/a=c/b=-2 \frac{k^2}{\sinh^2(k t)}, K_a=12 k^2, K_b=0,
}\\ \displaystyle{
a^2=0, b/a=    -  \frac{k^2}{\sinh^2(k t)}, c/b=0, K_b^2=4(K_a-3 k^2)(K_a-12 k^2)^2,
}\\ \displaystyle{
a^2=0, c/b=    -  \frac{k^2}{\sinh^2(k t)}, b/a=0, K_b^2=4(K_a-3 k^2)(K_a-12 k^2)^2,
}                                          
\end{array}
\right.
\label{eqa20Two-trigo-sol}
\end{eqnarray} 
but our goal is to obtain the four-parameter general solution.

The only method we could find to integrate this third degree ODE (\ref{eqode2fba})
is the one-family truncation method of Weiss \textit{et al.} \cite{WTC},
whose application is detailed in the Appendix.
One thus obtains the general solution of (\ref{eqode2fba}) 
\textit{via} the representation
\begin{eqnarray}
& & {\hskip -5.0truemm}
\left\lbrace
\begin{array}{ll}
    \displaystyle{b/a=-\chi^{-2} + \frac{  K_a}{3}  - \frac{3}{4} f^2 - f', \chi' = 1 + \frac{S(t)}{2} \chi^2(t),
}\\ \displaystyle{  S=-\frac{2 K_a}{3}  + \frac{3}{2} f^2 +3 f',
}\\ \displaystyle{f''  + 3 f f' + f^3- \frac{K_a}{3} f - \frac{K_b}{27} =0.
}                                          
\end{array}
\right.
\label{eqa0general}
\end{eqnarray}
The two nonlinear ODEs (for $\chi$ and $f$) are linearizable,
\begin{eqnarray}
& & {\hskip -5.0truemm}
\left\lbrace
\begin{array}{ll}
    \displaystyle{f=\frac{\varphi'}{\varphi}, \varphi^{(3)} - \frac{K_a}{3} \varphi' - \frac{K_b}{27} \varphi=0,
}\\ \displaystyle{\varphi=\sum_{n=1}^3 A_n e^{k_n t}, \sum k_n=0, 
K_a=-3 (k_2 k_3+k_3 k_1+k_1 k_2), K_b=27 k_1 k_2 k_3,
}\\ \displaystyle{\chi=\frac{\psi}{\psi'}, \psi'' + \frac{S}{2} \psi=0,
}\\ \displaystyle{\psi=\varphi^{-1/2} \left[
  k_1(k_2-k_3) A_1^{-1} e^{-k_1 t}
 +k_2(k_3-k_1) A_2^{-1} e^{-k_2 t}
 +k_3(k_1-k_2) A_3^{-1} e^{-k_3 t}\right],
}                                          
\end{array}
\right.
\end{eqnarray}
therefore $b/a$ and $c/b$ are rational expressions of two different exponential functions
$e^{k_j t}, j=1,2,3, k_1+k_2+k_3=0$.

\section{Conclusion}

About the search for the missing four-parameter solution \cite{LMC1994} of the mixmaster model,
since none of the above three cases of (\ref{System-natural})
includes an elliptic solution,
the present study unfortunately does not provide a hint to this missing solution.

On the other hand,
since the two six-dimensional Bianchi IX dynamical systems 
(\ref{System-natural}) and (\ref{eqBianchiIXMixmaster})
share two particular cases with the same kinds of particular singlevalued solutions,
we conjecture that the mixmaster system (\ref{eqBianchiIXMixmaster})
admits a third particular case whose solution would be similar to that of the case $a^2=0$
of (\ref{System-natural}), i.e.~rational in two exponentials.

\textit{Remark}.
One might wonder whether there exists some connection between the four-parameter solutions here found
and the general solution of interest to general relativity.
The answer is negative.
Indeed, let us consider the sixth order ODE $E_6(a^2)=0$ for $a^2(t)$, 
resulting from the elimination of the two other components,
and, for instance, the fourth order ODE $E_4(a^2)=0$ resulting from the elimination of $b/a$ between 
the system (\ref{System-natural-fcb-zero}) when $c/b=0$.
The general solution of $E_6$ has a maximal singlevalued subset depending on only two parameters,
as proven by the four irrational Fuchs indices (\ref{eqFuchs}),
while the general four-parameter solution of $E_4$ is singlevalued.
It is therefore impossible that $E_4$ be part of the general solution 
of interest to general relativity.
The equation $E_6=0$,
\begin{eqnarray}
& &  
E_6 \equiv (a^2)^{(6)} a^{10} E_4 + \hbox{polynomial}\left(\left\lbrace (a^2)^{(j)},j=0,\dots,5 \right\rbrace\right)=0,
\end{eqnarray}
is of the same nature as
\begin{eqnarray}
& &  
u u''-3{u'}^2=0,
\label{eqsimple}
\end{eqnarray}
i.e.~:
multivalued general solution $u=1/\sqrt{c_1 t + c_2}$,
singlevalued particular solution $u=1/\sqrt{c_2}$,
while the singlevalued expression $u=0$ is \textit{not} part of the general solution.

\vfill\eject
\begin{appendix}
\section*{Appendix. Integration of the case $a^2=0$}

In the one-family truncation method \cite{WTC}, 
one describes a movable singularity $t=t_0$ by a function $\chi(t)$
vanishing like $t-t_0$ and 
defined by \cite{Conte1989}
\begin{eqnarray}
& &  
\frac{\D}{\D t} \chi(t) = 1 + \frac{S(t)}{2} \chi^2(t),
\end{eqnarray}
in which $S(t)$ is a function to be found.
Then one assumes $b/a$ equal to the principal part of a Laurent series of $\chi(t)$
which matches the singularity structure (here a double pole),
\begin{eqnarray}
& &  
b/a=k_0\chi^{-2} + f_1(t) \chi^{-1} + f_2(t).
\end{eqnarray} 
Under these assumptions, the ODE (\ref{eqode2fba}) evaluates,
after elimination of all the derivatives of $\chi$,
to a similar principal part of a Laurent series of $\chi$, depending on $k_0, f_1(t), f_2(t), S(t)$,
which is required to identically vanish,
\begin{eqnarray}
& &  
(\ref{eqode2fba})= \chi^{-12} \sum_{j=0}^{12} E_j \chi^j, \forall j:\ E_j(k_0,f_1(t),f_2(t),S(t))=0.
\end{eqnarray} 
This set of 13 nonlinear ODEs is best solved by ascending values of $j$.
Equation $j=0$ factorizes as $(k_0+2)(k_0+1)=0$.
The value $k_0=-2$ yields successively $f_1=0$, then $f_2=K_a/18-(2/3) S$,
then $S'=0$, $K_b=0$, $K_a=-6 S$,
which only defines the particular solution
\begin{eqnarray}
& &  
\chi'=1-\frac{K_a}{12} \chi^2, 
b/a=c/b=-2 \frac{k^2}{\sinh^2(k t)}, 
k^2=\frac{K_a}{12}\cdot
\label{eqa20sol1}
\end{eqnarray} 

The second value $k_0=-1$ also yields $f_1=0$, then the next nonzero equation
\begin{eqnarray}
& &  
E_6 \equiv \left[K_b+27 S'+108 f_2' \right]^2-\left[K_a-12 S-36 f_2 \right]^2\left[K_a-3 S-9 f_2 \right]=0,
\end{eqnarray}
is solved for $f_2$ and $S'$ in terms of an auxiliary function $f(t)$,
\begin{eqnarray}
& & {\hskip -5.0truemm}
\left\lbrace
\begin{array}{ll}
\displaystyle{
f_2=\frac{K_a}{9} - \frac{S(t)}{3}-\frac{f^2(t)}{4},
}\\ \displaystyle{
S'=\frac{K_b}{9}  + K_a f(t) - 3 f^3(t) -6 f(t) f'(t).
}                                          
\end{array}
\right.
\end{eqnarray}

The next equation $j=7$ splits the resolution into three cases,
\begin{eqnarray}
& & {\hskip -10.0truemm}
f \left(f^2- \frac{K_a}{3}\right)
\left[
f f'' + {f'}^2 + 4 f^2 f' - \frac{4 K_a}{9} f' + \frac{5}{4} f^4- \frac{5 K_a}{9} f^2
      - \frac{K_b}{27} f - \frac{S^2}{9} + \frac{4}{81} K_a^2\right]=0.
\end{eqnarray}
The first factor $f=0$ yields the particular solution
\begin{eqnarray}
& &
f=0:
\begin{array}{ll}
\displaystyle{ 
K_b=0,\chi'=1-\frac{K_a}{3}\chi^2, \frac{b}{a}=-\frac{k^2}{\sinh^2(k t)}, \frac{c}{b}=0, k^2=\frac{K_a}{3}\ccomma
}
\end{array}
\label{eqa20sol23}
\end{eqnarray}
the second factor $f^2 - K_a/3=0$ defines again the same solution (\ref{eqa20sol23}),
\begin{eqnarray}
& &  {\hskip -10.0truemm}
f^2=-\frac{K_a}{3} \not=0:
K_b=0, \chi'=1-\frac{K_a}{12}\chi^2, \frac{b}{a}=-\frac{k^2}{\sinh^2(k t)}, \frac{c}{b}=0, k^2=\frac{K_a}{12}.
\end{eqnarray}
After solving the third factor for the pivot $f''(t)$,
the remaining equations $j=8:12$ have the common factor 
$18 f' + 9 f^2 - 6 S -4 K_a$, 
and this factor defines the general solution (\ref{eqa0general}).
After dividing the equations $j=8:12$
by the above-mentioned common factor,
one only obtains its particular solution $K_b^2=4 K_a^3$,
in which case $b/a$ and $c/b$ are rational functions of $t$ and $e^{k t}$.
The further degeneracy $K_b=K_a=0$ yields the rational 
particular solution,
\begin{eqnarray}
& & {\hskip -10.0 truemm}
K_b=K_a=0, 
a^2=0,
\frac{b}{a}=-2 \frac{t^2+c_1}{(t^2-c_1)^2}, 
\frac{c}{b}=-2 \frac{t^2-c_1}{(t^2+c_1)^2}, c_1 \hbox{ arbitrary}.
\end{eqnarray} 

\end{appendix}

\vfill\eject


\vfill\eject
\end{document}